\documentclass{ws-mpla}

\usepackage{cite}
\usepackage{xspace}
\usepackage{url}
\usepackage{units}

\begin{document}

\markboth{Alec Habig}
{MINOS neutrino oscillation results}

\catchline{}{}{}{}{}

\title{A Brief Review of MINOS neutrino oscillation results}

\author{\footnotesize ALEC HABIG}

\address{Physics Department, University of Minnesota Duluth, 10
  University Dr.\\
Duluth, MN 55812,
USA\\
ahabig@umn.edu}

\maketitle

\pub{Received (Day Month Year)}{Revised (Day Month Year)}

\begin{abstract}
  The MINOS long-baseline experiment is using the NuMI neutrino beam to
  make precise measurements of neutrino flavor oscillations in the
  ``atmospheric'' neutrino sector.  MINOS observes the $\nu_\mu$
  disappearance oscillations seen in atmospheric neutrinos, tests
  possible disappearance to sterile $\nu$ by measuring the neutral
  current flux, and extends our reach towards the so far unseen
  $\theta_{13}$ by looking for $\nu_e$ appearance in this $\nu_\mu$
  beam.  The magnetized MINOS detectors also allow tests of CPT
  conservation by discriminating between neutrinos and anti-neutrinos on
  an event-by-event basis.  The intense, well-understood NuMI neutrino
  beam created at Fermilab is observed \unit{735}{km} away at the Soudan
  Mine in Northeast Minnesota.  High-statistics studies of the neutrino
  interactions themselves and the cosmic rays seen by the MINOS
  detectors have also been made.  MINOS started taking beam data in May
  of 2005 and is now nearing the end of it's five-year run.  This paper
  reviews results published based on the first several years of data.

\end{abstract}

\ccode{PACS Nos.: 14.60.Pq, 14.60.Lm, 14.60.St, 29.27.-a, 29.30.-h}

\section{Introduction}	
\label{sec:intro}

One possible implication of non-zero neutrino mass is that as neutrinos
propagate, mixing occurs between mass eigenstates ($\nu_1, \nu_2,
\nu_3$) and the flavor eigenstates ($\nu_e, \nu_\mu, \nu_\tau$) involved
in the weak interactions governing neutrino production and detection.  The
mass and flavor eigenstates are related by a unitary matrix $U_{PMNS}$
expressed in terms of three mixing angles ($\theta_{12}, \theta_{13},
\theta_{23}$) and a $CP$-violating phase
$\delta$\cite{Maki:1962mu,Pontecorvo:1967fh}.  Experimental evidence
points towards this occurring via disappearance of $\nu_e$ in solar
neutrinos\cite{Hosaka:2005um,Aharmim:2008kc} and $\bar{\nu}_e$ in
reactor neutrinos\cite{:2008ee}, as well as $\nu_\mu$ disappearance in
atmospheric neutrinos produced by cosmic
rays\cite{Ashie:2004mr,Ashie:2005ik, Ambrosio:2004ig} and
accelerators\cite{Ahn:2006zza}.

The goal of the \underline{M}ain \underline{I}njector
\underline{N}eutrino \underline{O}scillation \underline{S}earch
(``MINOS'') long-baseline experiment is to precisely measure the
oscillation parameters involved in the atmospheric-sector oscillations.
It does this by observing the intense and well-understood NuMI beam over
a known baseline using with two similar magnetized steel/scintillator
calorimeters: \unit[1]{km} from its source at Fermilab with a
\unit[0.98]{kton} ``near detector'', then again \unit[735]{km} to the
northwest in the Soudan Mine Underground Lab using the \unit[5.4]{kton}
``far detector''.  Such a before-and-after comparison of the neutrinos
greatly reduces systematic errors associated with comparing differences
in the observed neutrino spectra to various neutrino oscillation
scenarios, allowing for a more accurate probe of the physics of neutrino
propagation.  Details of the detector designs, calibrations, and
performance can be found in \cite{Michael:2008bc}.

The NuMI beam\cite{Anderson:1998zz} is generated by \unit[120]{GeV}
protons hitting a carbon target.  The resulting charged pions are
focused by two electromagnetic horns and sent down a \unit[675]{m} decay
pipe, producing a beam of 92.9\%~$\nu_\mu$,
5.8\%~$\overline{\nu}_{\mu}$, 1.2\%~$\nu_e$ and
0.1\%~$\overline{\nu}_{e}$.  Changing the pion-focusing horn positions
and currents creates very different neutrino spectra.  The bulk of the
data come from the ``low energy'' beam configuration, peaked at several
GeV (see the dashed line in Fig.~\ref{fig:numuspectrum}).  Short exposures
in other beam configurations observed by the high-statistics near
detector provide good crosschecks to the beam modeling process and also
help to reduce systematic errors.

This paper summarizes the results of several analyses of the neutrino
data acquired over the two year time period starting with the beginning
of NuMI operations in May of 2005 and ending during the summer shutdown
in June 2007, an integrated exposure of over $3\times10^{20}$~protons on
target (``pot'') with a neutrino yield on order of one neutrino per
proton.  The intrinsic divergence of the beam results in a neutrino flux
at the far detector which is a factor of $10^6$ lower than that at the
near detector.  An additional \unit[$4\times10^{20}$]{pot} is on
tape, representing beam from October 2007 through June 2009.  The beam
focusing was reversed to favor anti-neutrino production upon resumption
of data taking in September of 2009, and was switched back to normal
neutrino mode in March 2010 after collecting
\unit[$1.8\times10^{20}$]{pot}.

MINOS analyses are done ``blind''.  That is, small subsets of the data
are examined immediately to develop the algorithms and cuts which
extract the physics results from the data, but the bulk of the dataset
is left alone till the data analysis methods have been settled on.
Detailed comparisons of variables used in cuts are made between
simulated Monte Carlo data and real data, to be sure that detector
quirks and parameter distributions are well modeled and understood.
Once the simulations and analysis techniques are close to final and
uncertainties, systematic errors and expected sensitivities calculated,
``sidebands'' (parameters similar to but not actually the ones used to
produce the final answer) are revealed and compared to expectations.
Only once everything is as well understood as possible is the ``box
opened'' and the mature analysis unleashed on the full dataset to
produce the final physics results.  This process helps to ensure that
the development of the analysis doesn't inadvertently (consciously or
otherwise) converge on a fluctuation in the data, skewing the final
results or their significances.  Extensions to the analyses presented
below to the newer, as yet unanalyzed data are in progress, and as new
improvements are being made to the methods, the new analyses are also
using this blinding procedure.

\section{Oscillation Analyses}
\label{sec:analyses}

MINOS was intended to observe Charged Current (``CC'') quasi-elastic
$\nu_\mu$ interactions to make a precision measurement of $\Delta
m^2_{32}$.  Its calorimeters are designed to measure the $\mu$ produced
in $\nu_\mu$ CC interactions, and are magnetized to provide momentum and
charge discrimination on an event by event basis, as well as the
standard calorimetric method of establishing particle energies via
$dE/dx$.  By using far and near detectors which are as similar as
possible, uncertainties in the energy scale between measurements of the
neutrino spectrum before and after traveling the \unit[735]{km} baseline
are minimized.  Small differences between detectors in light collection
and electronics were cross-calibrated in a beam test at CERN using the
``calibration detector''\cite{Adamson:2006xv} which observed the same
particles with both sets of electronics\cite{Cabrera:2009fi}.

Since the oscillation minima at this baseline is less than the $\tau$
production threshold energy, the oscillatory signature is that of
$\nu_\mu$ disappearance rather than $\nu_\tau$ appearance, an analysis
described in Sec.~\ref{sec:CC}.  MINOS' magnetic fields allow
discrimination between $\nu_\mu$ and $\bar{\nu}_\mu$ and provide a
chance to test CPT conservation by seeing if neutrinos and
anti-neutrinos have the same oscillation parameters
(Sec.~\ref{sec:numubar}).  Although the calorimetry is coarser than one
might like for such a measurement, electromagnetic showers can still be
resolved.  This allows further study of possible $\nu_\mu$ disappearance
to sterile $\nu$ states by observation of Neutral Current (``NC'')
interactions (Sec.~\ref{sec:sterile}), and sensitivity to sub-dominant
$\nu_\mu\leftrightarrow\nu_e$ transitions by looking for $\nu_e$
appearance oscillations (Sec.\ref{sec:nue}).

\subsection{$\nu_\mu$ Disappearance Oscillations}
\label{sec:CC}

Since $\theta_{23}\gg\theta_{13}$ and $\theta_{12}$, when considering
the atmospheric neutrino sector the full neutrino mixing matrices reduce
to a simpler two-flavor equation.  A $\nu_\mu$ of energy
$E_\nu[\rm{GeV}]$ observed after traveling some distance $L[\rm{km}]$
from its production point has a probability of being detected as a
$\nu_\mu$ given by
\begin{equation}
P(\nu_{\mu} \rightarrow \nu_{\mu})=1-\sin^2(2\theta_{23})
  \sin^2\left(1.27\Delta m_{32}^2\frac{L}{E}\right),
  \label{eq:2flavor}
\end{equation}
where $\Delta m^2[eV^2]$ is the mass difference between $\nu_2$ and
$\nu_3$ and $\sin^2(2\theta)$ is the mixing amplitude.  

\begin{figure}[t!]
\centerline{\psfig{file=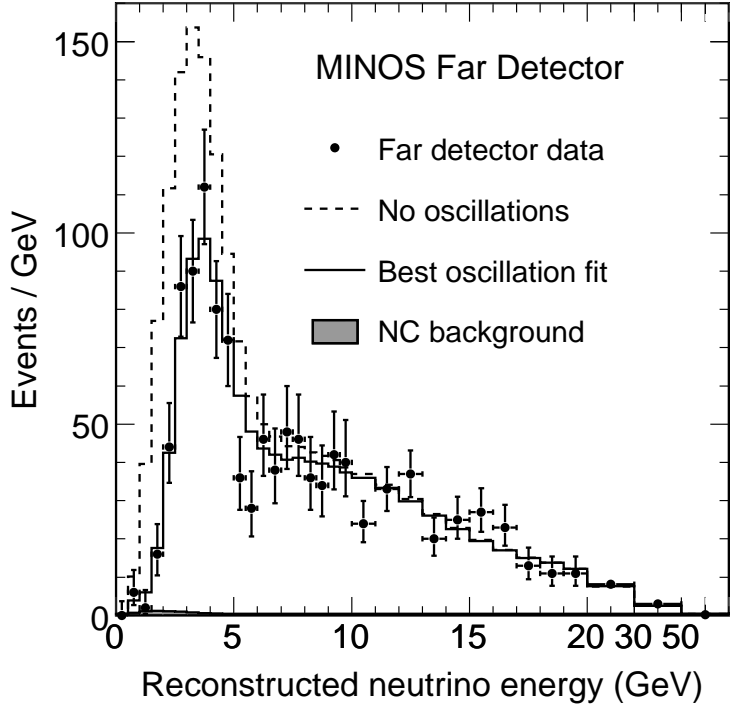,width=0.9\textwidth}} 
\caption{\label{fig:numuspectrum}The MINOS far detector $\nu_\mu$
  spectrum\cite{Adamson:2008zt}.  The data (points with statistical
  errors) show a significant deficit from the null hypothesis (dashed
  line),
  but well-match a $\nu_\mu\leftrightarrow\nu_\tau$ oscillation scenario
  (solid line), with best fit mass splitting $|\Delta m^ 2|=(2.43\pm
  0.13)\times10^{-3}$~eV$^2$ (68\% cl) and mixing angle
  $\sin^2(2\theta)>0.90$~(90\% cl).}
\end{figure}

An exposure of \unit[$3.36\times10^{20}$]{pot} in MINOS has been
analyzed\cite{Adamson:2008zt}, selecting 848 far detector events as
$\nu_\mu$ with good purity.  The observed (unoscillated) near detector
signal is used to calculate a null hypothesis expectation of 1065$\pm$60
far detector events, including a small background estimated to be
composed of 2.3 external $\mu$, 5.9 NC induced showers, and 1.5 $\tau$
decays.  The resulting spectra is shown with the observed data in
Fig.~\ref{fig:numuspectrum}.  Comparing the observed data to
expectations modified by Eq.~\ref{eq:2flavor} result in best fit
oscillation parameters of $|\Delta m^
2|=(2.43\pm0.13)\times10^{-3}$~eV$^2$ (68\% cl) and
$\sin^2(2\theta)>0.90$~(90\% cl).  Systematic errors in $\Delta m^2$ are
dominated by uncertainties in the hadronic energy scale ($\pm$10.3\%
absolute and $\pm$3.3\% relative between near and far detectors) and the
relative normalization between detectors ($\pm$4\%).  The background of
NC showers mis-reconstructed as $\nu_\mu$ quasi-elastic interactions
($\pm$50\%) dominates in $\sin^22\theta_{23}$.  These systematics
(\unit[$\pm0.108\times10^{-3}$]{eV$^2$} in $\Delta m^2$ and $\pm0.018$
in $\sin^22\theta_{23}$) are still smaller than the statistical
(\unit[$\pm0.19\times10^{-3}$]{eV$^2$} in $\Delta m^2$ and $\pm0.09$ in
$\sin^22\theta_{23}$) errors, so the measurement will improve as data
from the remaining $\sim\frac{2}{3}$ of the MINOS exposure are added.

\begin{figure}[h]
\centerline{\psfig{file=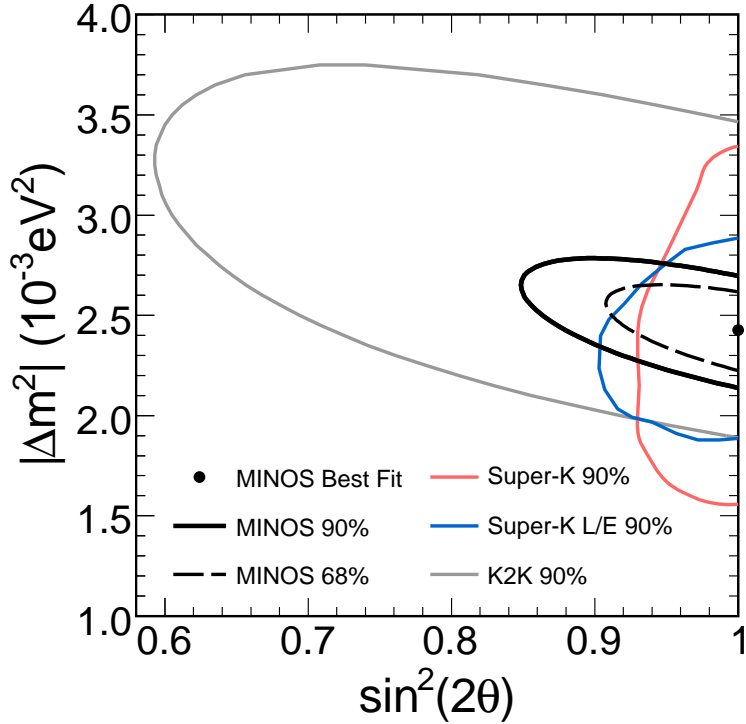,width=0.9\textwidth}} 
\caption{\label{fig:allowed}The resulting allowed region in oscillation
  parameter space for the data in Fig.~\ref{fig:numuspectrum}, at 90\%
  (solid black) and 68\% (dashed black) confidence levels.  Two Super-K
  atmospheric analyses are shown in red\cite{Ashie:2005ik} and
  blue\cite{Ashie:2004mr}, and the K2K long baseline experiment results
  produce the gray contour\cite{Ahn:2006zza}.}
\end{figure}

\subsection{Anti-neutrinos}
\label{sec:numubar}

The magnetized nature of the MINOS detectors allows the event-by-event
determination of the charge sign of muons, and thus the identification
of the parent neutrino or anti-neutrino undergoing the quasi-elastic
interaction that produced them.  Selection of wrong-sign muons in the
$\nu_\mu$ beam tests if $\bar{\nu}_\mu$ oscillate in the same fashion as
$\nu_\mu$ in Eq.~\ref{eq:2flavor}.  Is $\bar{\theta}_{23}=\theta_{23}$
and $\Delta\bar{m}^2_{32}=\Delta m^2_{32}$?  Furthermore, could some
fraction $\alpha$ of disappearing $\nu_\mu$ reappear as $\bar{\nu}_\mu$?
These are both tests of CPT conservation, and with its magnetic field
MINOS is the only experiment capable of testing this.  

\begin{figure}[h]
\centerline{\psfig{file=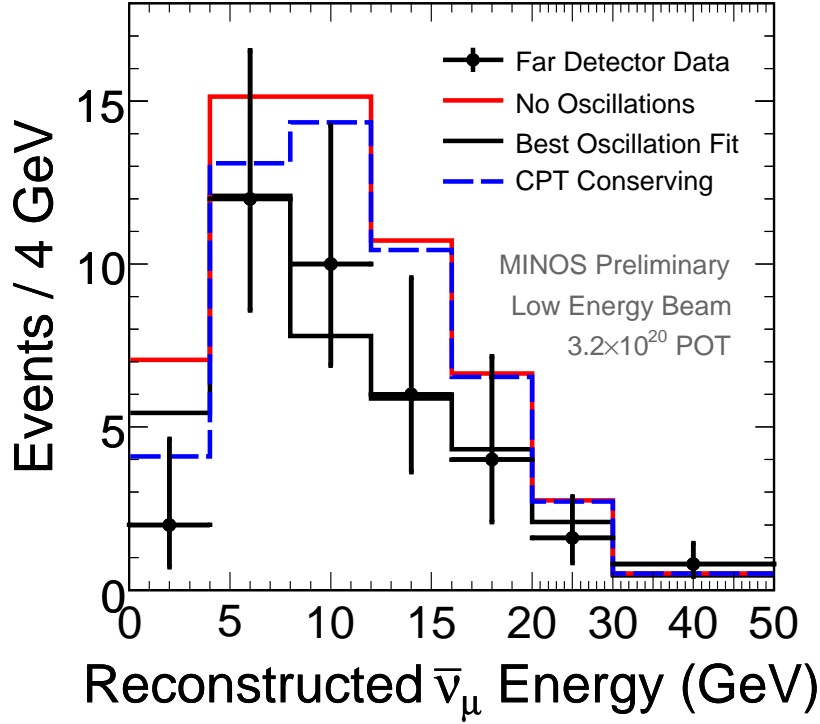,width=0.9\textwidth}} 
\caption{\label{fig:numubar}The MINOS far detector $\bar{\nu}_\mu$
  spectrum\cite{Himmel:2009zz}.  The data (points with statistical
  errors) are consistent with CPT-conserving neutrino oscillations (blue
  dashed line) within the large error bars.}
\end{figure}

Anti-neutrinos in the NuMI beam come primarily from $\pi^+$ which travel
directly down the center of the focusing horns and thus avoid being
de-focused, leaving little kinematic phase space for these pions.
Combined with the lower cross-sections for anti-neutrinos compared to
neutrinos, only 6.4\% of the neutrino interactions in a
\unit[$3.2\times10^{20}$]{pot} far detector exposure are due to
anti-neutrinos, making the relative backgrounds are higher, the
statistics lower, and the spectrum harder (Fig.~\ref{fig:numubar}).
42~$\bar{\nu}_\mu$ events are seen while
$64.6\pm8.0_{stat}\pm3.9_{syst}$ are expected in the no-oscillation
case, or $58.3\pm7.6_{stat}\pm3.6_{syst}$ if CPT is conserved given the
observed $\nu_\mu$ oscillation parameters\cite{Himmel:2009zz}.  This
places a 90\%cl upper limit of $\alpha<0.026$, and the anti-neutrino
oscillation parameters are consistent with the neutrino parameters given
these low statistics, excluding \unit[$5.0<\Delta\bar{m}^2_{32} <
81$]{eV$^2$} at 90\%cl for maximal mixing.

\begin{figure}[h]
\centerline{\psfig{file=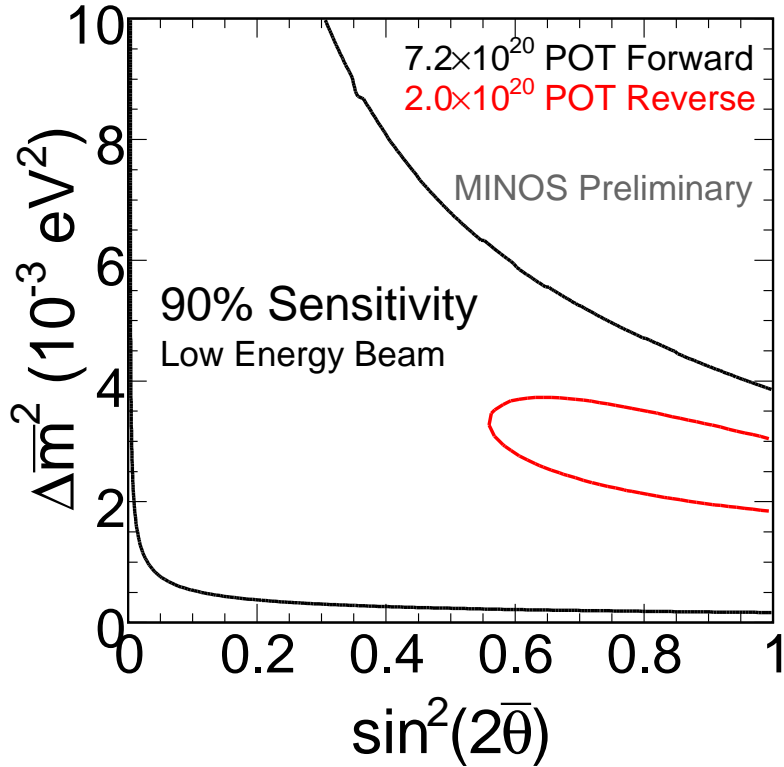,width=0.9\textwidth}} 
\caption{\label{fig:rhc} Sensitivity contours comparing the oscillation
  parameters available using just the anti-neutrinos from the neutrino
  beam, as described in the text but including all available data
  (\unit[$7.2\times10^{20}$]{pot}, the black line) to that obtainable
  using a \unit[$7.2\times10^{20}$]{pot} exposure of dedicated
  anti-neutrino beam (red contour).  This red contour is comparable in
  scope to that of the K2K experiment's neutrino results (the gray line
  in Fig.~\ref{fig:allowed}).}
\end{figure}

To increase the anti-neutrino dataset by an order of magnitude, the
polarity of the focusing horns in the NuMI beam was reversed to focus
$\pi^+$ rather than the usual $\pi^-$ for an exposure of
\unit[$1.8\times10^{20}$]{pot} of anti-neutrino production.  The
statistics available in this set of data would make a reasonable
measurement of $\bar{\theta}_{23}$ and $\Delta\bar{m}^2_{32}$
(Fig.\ref{fig:rhc}).

\subsection{Sterile Neutrinos}
\label{sec:sterile}

Another possible explanation of $\nu_\mu$ disappearance is oscillation
into sterile neutrinos (``$\nu_s$'') which experience no interactions
and thus would disappear.  This would also suppress the rate of NC
events in the far detector compared to the traditional explanation of
oscillation to sub-threshold $\nu_\tau$, since those $\nu_\tau$ still
undergo NC interactions but $\nu_s$ would not.  

\begin{figure}[h]
\centerline{\psfig{file=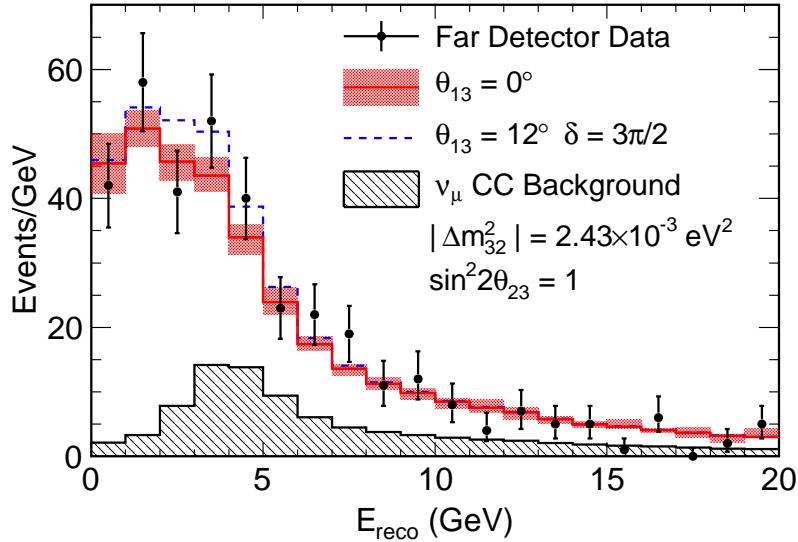,width=0.9\textwidth}} 
\caption{\label{fig:ncspectrum}The MINOS far detector neutral current
  spectrum\cite{Adamson:2010wi}.  The data (points with statistical
  errors) track the predicted spectrum (red hashed boxes) and do not
  show the deficit at lower energies expected from an scenario which
  involves the NuMI $\nu_\mu$ changing to something which does not
  undergo such interactions, either a sterile neutrino or by decaying
  away entirely.  The dashed line is the prediction for showers should
  there be $\nu_e$ appearance at the CHOOZ limit, as they would be
  included in this data selection.  The gray shaded region is the
  expected background, primarily composed of mis-reconstructed charged
  current events.}
\end{figure}

To test this hypothesis, NC showers have been selected from an exposure
of \unit[$3.18\times10^{20}$]{pot}\cite{Adamson:2010wi}.  A NC
interaction produces no outgoing leptons, but simply a hadronic shower.
Any $\pi^0$ produced in that shower decays rapidly to a pair of gamma
rays, which produce diffuse electromagnetic showers that can be reliably
separated from the long $\mu$ tracks used in Sec.~\ref{sec:CC}.  388
events are selected, 141 of them of less than \unit[3]{GeV}, with an
estimated 17 non-NC interactions (primarily very short track $\nu_\mu$
CC interactions) creeping in as background in this low energy region of
interest for oscillation physics.  The resulting spectrum of these NC
events (Fig.~\ref{fig:ncspectrum}) is not depressed compared with the
expectations of NC interactions from standard
$\nu_\mu\leftrightarrow\nu_\tau$ oscillations.  The ratio of observed to
expected NC events in the far detector is
$R=0.99\pm0.09_{stat}\pm0.07_{syst}-0.08_{\nu_e}$ in the energy region
of \unit[0-3]{GeV} where $\nu_\mu$ are disappearing.  While the addition
of a fourth neutrino eigenstate results in several new ways things could
oscillate, fits to all those models place limits on $\theta_{24}$ and
$\theta_{34}$ on being half (or less) than $\theta_{23}$, showing that
sterile neutrinos are not a dominant player in $\nu_\mu$ disappearance.
Another way to state this is as a limit on the fraction of $\nu_s$
participation of $f_s<0.51$ at 90\%cl.

\subsection{The Search for $\nu_e$ Appearance}
\label{sec:nue}

The CHOOZ experiment\cite{Apollonio:1999ae} sets an upper limit of
$\sin^{2}(2\theta_{13})<0.15$ on the mixing amplitude governing the
transmutation of NuMI $\nu_\mu$ into $\nu_e$.  MINOS was designed to be
a good muon calorimeter for $\nu_\mu$ disappearance, but is coarse for
resolution of $\sim$GeV electromagnetic showers, each \unit[2.54]{cm
  steel}+\unit[1.0]{cm plastic} thick plane being 1.4~radiation lengths
thick, and each \unit[4.1]{cm} wide scintillator strip being
1.1~Moli\`ere radii.  However, the experiment retains sensitivity to the
$\sim$2\% $\nu_e$ appearance signal which a $\theta_{13}$ near the CHOOZ
limit\cite{Apollonio:1999ae} would create, and the first
\unit[$3.14\times10^{20}$]{pot} of MINOS data have been
examined\cite{Collaboration:2009yc} by a neural network to select
electromagnetic shower candidates, which are more compact when produced
by an electron from a CC $\nu_e$ interaction than from a NC-induced
$\pi^0$.  When applied to Monte Carlo data this is 41\% efficient at
keeping $\nu_e$ events while rejecting $>$92\% of NC showers (the
dominant background) and $>$99\% of $\nu_\mu$ charged current (``CC'')
interactions (high-y collisions which put much more energy into the
hadronic debris than the outgoing $\mu$ lepton).

\begin{figure}[h]
\centerline{\psfig{file=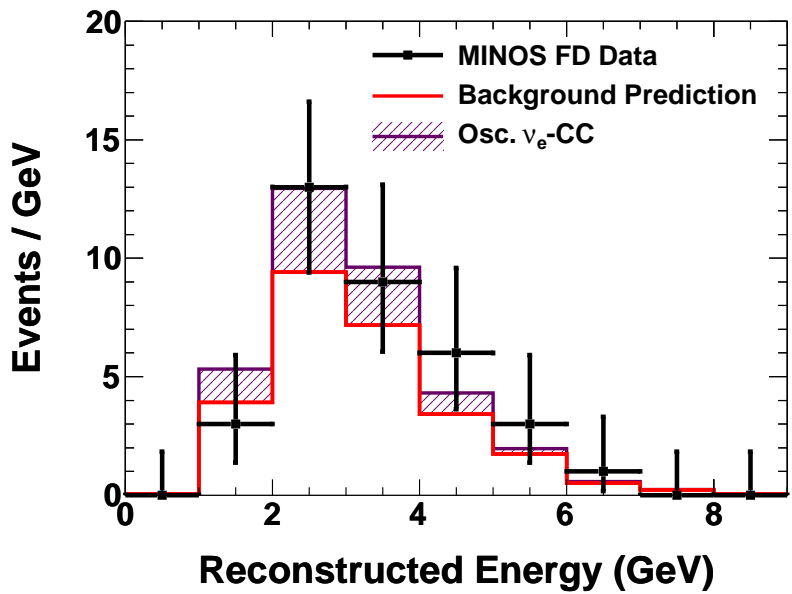,width=0.9\textwidth}} 
\caption{\label{fig:nue}The spectrum of potential $\nu_e$ interactions
  in the MINOS far detector, with statistical plus systematic error
  bars\cite{Collaboration:2009yc}.  The $1.5\sigma$ excess is consistent
  with both the expected large background (red) and a
  $\sin^{2}(2\theta_{13})$ comparable to the CHOOZ
  limit\cite{Apollonio:1999ae} (purple).}
\end{figure}

Given the small expected signal and large uncertainties in hadronic
shower modeling, data-driven methods are used to better estimate the
background.  At the near detector no oscillation has yet occurred, so
with the exception of the well-modeled inherent beam $\nu_e$, all events
selected must be examples of such background events.  The two classes of
backgrounds have different production kinematics so extrapolate to the
Far Detector slightly differently, thus two techniques are used to
deconvolve the background.  The first takes obvious $\nu_\mu$
interactions and subtracts the hits from the muon track, resulting in a
sample of CC-induced showering events to study.  The second compares
data from beam running with the focusing horn on or off.  The very
different neutrino spectra which result allow fitting for the two
background components.  Both methods give comparable results and produce
the background at the far detector shown as the red line in
Fig.~\ref{fig:nue}.  This yields an expected background of 26.6
(18.2~NC, 5.1~CC, and 2.2 beam $\nu_e$) at the far detector, while 35
$\nu_e$ like events are seen, a \unit[1.5]{$\sigma$} excess (including
7.3\% statistical and 19\% systematic errors) (Fig.~\ref{fig:nue}).  If
fit for oscillations, this is just below the CHOOZ limit and consistent
within errors with no $\nu_e$ appearance.

There is somewhat more than twice the exposure used for this analysis
already on tape.  In addition to decreasing the statistical error on the
data points, the systematic error bars include a large statistical
component due to the small data subsets used in the data-driven
background estimation.  Thus, an analysis of the additional data
available is expected to substantially reduce the systematic error bars
as well.  Projections of the sensitivity of this new dataset suggest
that the slight excess observed in this current analysis will either be
shown to be more significant (if it is really $\nu_e$ appearance) or to
be revealed as merely a background fluctuation.

\subsection{Alternate Hypotheses}
\label{sec:althyp}

\begin{figure}[h]
\centerline{\psfig{file=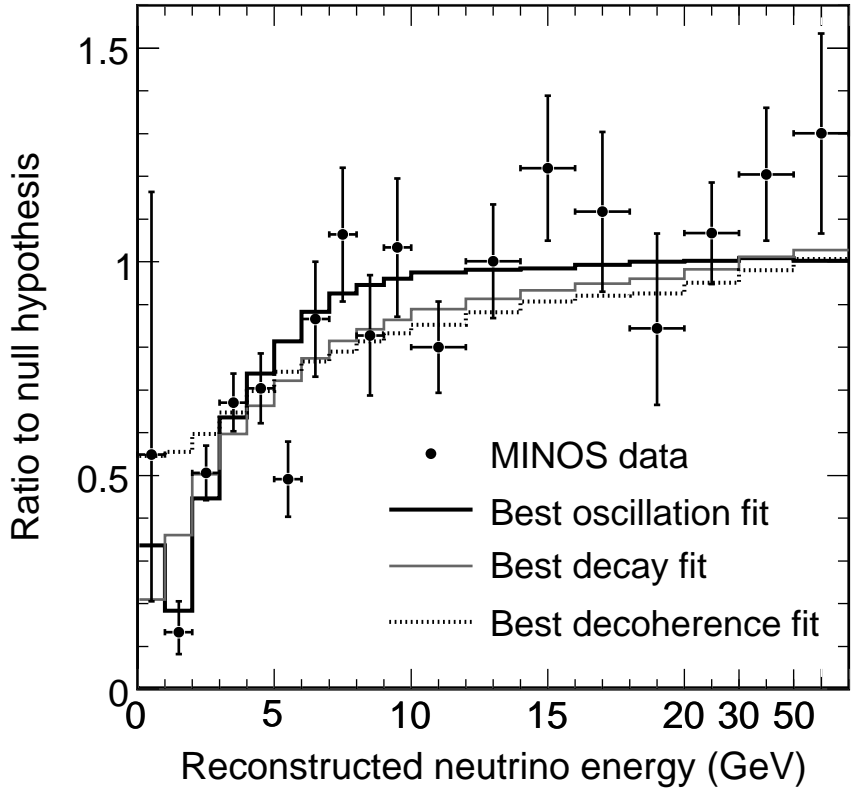,width=0.9\textwidth}} 
\caption{\label{fig:ccratio}The ratio of the MINOS far detector data
  from Fig.~\ref{fig:numuspectrum} to the no-oscillation null
  hypothesis, from \cite{Adamson:2008zt}.  Superimposed are the best fit
  expectations of three $\nu_\mu$ disappearance models: standard
  $\nu_\mu\leftrightarrow\nu_\tau$ (thick solid line), which fits the
  data the best; pure neutrino decay\cite{Barger:1998xk} (thin solid
  line), which is disfavored at the \unit[3.7]{$\sigma$} level; and
  quantum decoherence\cite{Fogli:2003th} (dashed line), which is
  disfavored by this data at \unit[5.7]{$\sigma$}.}
\end{figure}

Two other hypotheses have been presented (and not thoroughly ruled out
by previous observations) to explain the $\nu_\mu$ disappearance outside
the standard oscillations model.  The first is quantum decoherence of
the neutrino's wave packet\cite{Fogli:2003th}, in which the survival
probability equivalent to Eq.~\ref{eq:2flavor} varies as
$[1-\exp(\frac{m^2L}{2E})]$ (Eq.~5 of\cite{Fogli:2003th}) rather than
$\sin^2(1.27L/E)$.  The best fit to this function is shown in
Fig.~\ref{fig:ccratio} as the dashed line, and a comparison of $\chi^2$
to the data and the standard $\nu_\mu\leftrightarrow\nu_\tau$ disfavors
the decoherence hypothesis at the \unit[5.7]{$\sigma$} level.

\begin{figure}[h]
\centerline{\psfig{file=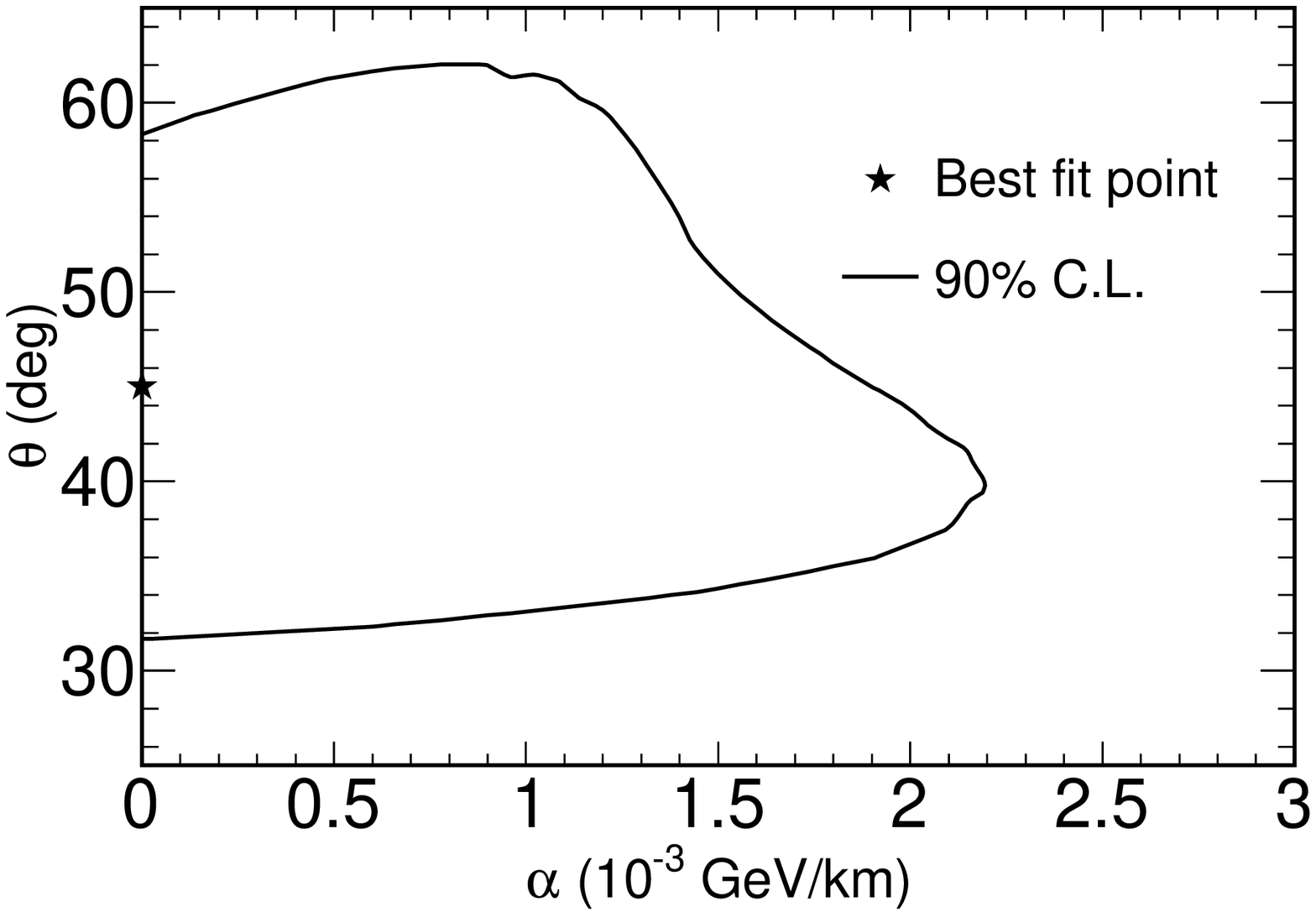,width=0.9\textwidth}} 
\caption{\label{fig:ncalpha}MINOS' 90\%cl allowed region for the
  neutrino mass/lifetime ration $\alpha$ compared to the oscillation
  mixing angle $\theta$\cite{Adamson:2010wi}.  The best-fit is for no
  decay (infinite lifetime), with an upper limit of
  \unit[$\tau_3/m_3 > 2.1\times10^{-12}$]{s/eV}.}
\end{figure}

The second alternate hypothesis is that neutrinos
decay\cite{Barger:1998xk}, producing a survival probability which varies
as $[\sin^2\theta+\cos^2\theta \exp(\frac{-\alpha L}{2E})]^2$ (Eq.~13 of
\cite{Barger:1998xk}).  Again, this shape does not match the data in
Fig.~\ref{fig:ccratio} as well as standard oscillations, and is
disfavored at \unit[3.7]{$\sigma$}.  However, this result is for pure
neutrino decay, which is also disfavored by Super-K\cite{Ashie:2004mr}.
If there were oscillations and neutrino
decay happening at the same time\cite{GonzalezGarcia:2008ru}, the waters
are muddier, see Eq.~17 of \cite{Adamson:2010wi}.  In order to get a
better handle on this problem, MINOS makes use of NC events as well as
the CC spectral shape\cite{Adamson:2010wi}, since a decayed neutrino
will not produce a NC event while a $\nu_\tau$ will.  Doing this
improves the rejection of the pure decay hypothesis to
\unit[5.4]{$\sigma$}, and places a limit on the ratio of neutrino mass
eigenstate $m_3$'s lifetime to mass of \unit[$\tau_3/m_3 >
2.1\times10^{-12}$]{s/eV} (Fig.~\ref{fig:ncalpha}).

\subsection{Atmospheric Neutrinos}
\label{sec:atmnu}

In addition to the NuMI beam, the MINOS far detector is bathed in the
same flux of cosmic-ray induced atmospheric neutrinos which provided the
first measurements of neutrino oscillations\cite{Fukuda:1998mi}.
However, MINOS being an order of magnitude smaller than
Super-Kamiokande, the atmospheric neutrino interaction rate is several
per week rather than dozens per day, limiting the statistical
significance of such measurements.  Nevertheless, analyses of such
neutrinos are consistent with the oscillation parameters established in
the beam neutrinos\cite{Adamson:2007vt} and provide the first direct
observations of anti-neutrinos from cosmic rays\cite{Adamson:2005qc}.

\section{Non-Oscillation analyses}
\label{sec:nonosc}

This review is of MINOS' neutrino oscillation results, so other work
using these detectors will not be discussed in depth.  However, studies
using cosmic rays have grown out of the need to calibrate the MINOS
detectors, and speak to the depth of understanding of these detectors.
Cosmic ray analyses include the first direct measurement of the charge
ratio of cosmic ray muons at TeV energies\cite{Adamson:2007ww} and probe
the meson production in cosmic ray primary interactions by watching the
variation in the underground muon rate vary with stratospheric
conditions\cite{Adamson:2009zf}, a competition between secondary mesons
decaying to produce the observed cosmic ray muon and re-interacting in
the atmosphere.  This effect is also of use to atmospheric physicists,
who turn the problem around to use the cosmic rays to study unusual
events in the stratosphere itself\cite{:2009zzh}.

Studies of neutrino interactions themselves are also a large topic of
work, as the near detector observes $O(10^4)$ neutrino interactions per
day of operations, by far the largest statistics sample in the world.
The resulting improved knowledge of neutrino interaction physics feeds
directly back to reducing the systematic errors in the oscillation
analyses\cite{Adamson:2009ju}.  Looking at MiniBOONE neutrinos in MINOS
and vice-versa helps to understand the off-axis components of neutrino
beams\cite{Adamson:2008qj}.  Using beam neutrinos in different ways has
also yielded interesting results on the velocity of neutrinos (comparing
arrival times at near and far detectors)\cite{Adamson:2007zzb} and has
been used to test for violation of Lorentz Invariance in the neutrino
sector\cite{:2008ij}.

\section{Conclusions}
\label{sec:conclusions}

MINOS has measured neutrino oscillation parameters in the
``atmospheric'' $\nu_2\leftrightarrow\nu_3$ sector with high precision,
favoring standard $\nu_\mu\leftrightarrow\nu_\tau$ oscillations with
$|\Delta m^2|=(2.43\pm0.13)\times10^{-3}$~eV$^2$ (68\% cl) and
$\sin^2(2\theta)>0.90$~(90\% cl).  Quantum decoherence as an explanation
for the $\nu_\mu$ disappearance is disfavored at the
\unit[5.7]{$\sigma$} level.  Measurements of the total active neutrino
flux using neutral current interactions help to disfavor pure
$\nu_\mu\leftrightarrow\nu_s$ oscillations by \unit[5.4]{$\sigma$},
place a limit on the ratio of neutrino mass eigenstate $m_3$'s lifetime
to mass of \unit[$\tau_3/m_3 > 2.1\times10^{-12}$]{s/eV}, and limit the
participation of sterile neutrinos in a sub-dominant mode to a fraction
$f_s<0.51$ at 90\%cl.  Examination of the limited set of anti-neutrino
data have shown no evidence for different oscillation parameters for
anti-neutrinos, and limit the fraction $\alpha$ of $\nu_\mu$
disappearing to $\bar{\nu}_\mu$ to $\alpha<0.026$ at 90\%cl.

These results come from an exposure of roughly 1/3 the eventual complete
MINOS dataset, so the final precision of the measurements will be
greater than those reviewed here.  This is especially true in the cases
of $\nu_e$ appearance, where the data-driven background estimations
methods will benefit greatly from additional statistics, and in the
measurements of $\bar{\nu}_\mu$, which will take advantage of dedicated
anti-neutrino beam running.

\section*{Acknowledgments}

The author would like to thank his many MINOS colleagues whose work is
being presented in this paper.  MINOS is supported by the U.S.
Department of Energy, the U.K. Science and Technologies Facilities
Council, the U.S. National Science Foundation, and the State and
University of Minnesota.  The author is supported by NSF RUI grant
\#0653016.


\end{document}